\documentclass[12pt]{article}

\usepackage[letterpaper,margin=1in]{geometry}
\usepackage{fontspec}
\usepackage{graphicx}
\usepackage{booktabs}
\usepackage{longtable}
\usepackage{array}
\usepackage{calc}
\usepackage{caption}
\usepackage{float}
\usepackage{amsmath}
\usepackage{amssymb}
\usepackage{xcolor}
\usepackage{setspace}
\usepackage{newunicodechar}
\usepackage{xurl}
\usepackage[hidelinks]{hyperref}

\graphicspath{{./}}

\providecommand{\textsubscript}[1]{\ensuremath{_{\mathrm{#1}}}}

\newunicodechar{₀}{\textsubscript{0}}
\newunicodechar{₁}{\textsubscript{1}}
\newunicodechar{₂}{\textsubscript{2}}
\newunicodechar{₃}{\textsubscript{3}}
\newunicodechar{₄}{\textsubscript{4}}
\newunicodechar{₅}{\textsubscript{5}}
\newunicodechar{₆}{\textsubscript{6}}
\newunicodechar{₇}{\textsubscript{7}}
\newunicodechar{₈}{\textsubscript{8}}
\newunicodechar{₉}{\textsubscript{9}}
\newunicodechar{⁰}{\textsuperscript{0}}
\newunicodechar{¹}{\textsuperscript{1}}
\newunicodechar{²}{\textsuperscript{2}}
\newunicodechar{³}{\textsuperscript{3}}
\newunicodechar{⁴}{\textsuperscript{4}}
\newunicodechar{⁵}{\textsuperscript{5}}
\newunicodechar{⁶}{\textsuperscript{6}}
\newunicodechar{⁷}{\textsuperscript{7}}
\newunicodechar{⁸}{\textsuperscript{8}}
\newunicodechar{⁹}{\textsuperscript{9}}
\newunicodechar{⁻}{\textsuperscript{-}}
\newunicodechar{⁺}{\textsuperscript{+}}
\newunicodechar{≥}{\ensuremath{\geq}}
\newunicodechar{≤}{\ensuremath{\leq}}
\newunicodechar{≈}{\ensuremath{\approx}}
\newunicodechar{×}{\ensuremath{\times}}
\newunicodechar{→}{\ensuremath{\rightarrow}}
\newunicodechar{−}{\ensuremath{-}}
\newunicodechar{°}{\ensuremath{^\circ}}
\newunicodechar{µ}{\ensuremath{\mu}}

\title{Energy-Limited Radiolytic Habitability in the Shallow Martian Subsurface: Implications for ExoMars Rosalind Franklin and Tianwen-3}
\author{Dimitra Atri\\
\small Space Exploration Laboratory\\
\small Center for Astrophysics and Space Science\\
\small New York University Abu Dhabi\\
\small \href{mailto:atri@nyu.edu}{atri@nyu.edu}}
\date{}

\begin{document}

\maketitle

\section{Abstract}\label{abstract}

The surface of Mars is sterilized by ionizing radiation and pervasive
oxidants; its shallow subsurface, shielded from ultraviolet light and
the most reactive oxidation, may instead preserve habitable conditions.
The radiolytic habitable zone (RHZ) hypothesis holds that galactic cosmic
rays can drive water and oxychlorine radiolysis there, generating H₂ and
oxidants that support chemolithotrophy without sunlight or geothermal
heat. We develop a quantitative framework coupling Monte Carlo GCR
transport, phase-resolved radiolysis chemistry, water-activity and
H₂-retention treatments, and microbial maintenance-power constraints for
Gale Crater, Oxia Planum, southern Utopia, and Arabia/Mawrth, expressing
radiolytic chemical energy as a depth-resolved redox power comparable to
the power requirements of life. H₂ retention is the dominant control.
Sorbed or mineral-associated retention yields 4--6 × 10⁻¹³ W kg⁻¹ at
protected depth (≥10 cm), whereas connected-pore free-gas escape lowers
redox power by two to four orders of magnitude. Gale remains inactive
under both retention assumptions, consistent with published SAM
evolved-gas analyses. Even in retained-H₂ active terrains, supported
cell density reaches only 10³--10⁵ cells cm⁻³ at subseafloor maintenance
powers. The framework predicts a spatially restricted, low-density RHZ,
testable by stepped H₂ evolved-gas analysis of protected-depth samples
from ExoMars Rosalind Franklin, Tianwen-3, or Mars Sample Return.

\textbf{Keywords:} Mars; radiolysis; radiolytic habitable zone; galactic
cosmic rays; subsurface habitability; hydrogen retention; supported cell
density; maintenance power; ExoMars Rosalind Franklin; Tianwen-3; Mars Sample
Return.

\section{1. Introduction}\label{introduction}

The shallow martian subsurface has been proposed as a candidate
environment for extant or recently extinct microbial life, including
radiolysis-supported chemolithotrophy in the radiolytic habitable zone
(Atri, 2016, 2020; Davila and Schulze-Makuch, 2016). Below the
ultraviolet-sterilized and oxidant-saturated near surface, reduced
chemical species and transient liquid or film-like water phases may
persist in stratigraphic intervals less affected by surface radiation
and oxidant cycling (Pavlov et al., 2012; Lasne et al., 2016). Those
intervals overlap the sampling horizon of landed and near-future
shallow-subsurface investigations, including Curiosity drilled samples,
ExoMars Rosalind Franklin's 2 m drill, and proposed sample-return contexts
(Sutter et al., 2017; Quantin-Nataf et al., 2021; Hou et al., 2025). The
central question is whether abiotic energy production at these depths
can be sustained at levels relevant to microbial maintenance.

Galactic cosmic rays (GCRs) can provide such a source through
radiolysis. GCRs penetrate the upper several meters of regolith and
ionize water- and oxychlorine-bearing phases, producing molecular
hydrogen as a reduced donor together with H₂O₂, O₂, and chlorate as
oxidant or oxidant-precursor species (Dartnell et al., 2007; Quinn et
al., 2013). Where these products coexist with
accessible electron acceptors, the resulting redox disequilibrium can be
harnessed by chemolithotrophic metabolism (Hoehler and Jørgensen, 2013;
LaRowe and Amend, 2015). This mechanism was proposed as the basis for a
radiolytic habitable zone (RHZ) on Mars by Atri (2016, 2020). Atri et
al.~(2025) subsequently placed the hypothesis on a quantitative footing
by evaluating GCR-driven radiolysis as a possible energy source for life
on Mars in the absence of photosynthesis or accessible geothermal heat.

This study develops a quantitative, physically explicit framework for
the RHZ hypothesis and evaluates its predictions in mission-accessible
martian terrains. The RHZ hypothesis already predicts
that radiolysis can supply energy in the shallow subsurface, so we focus
on two further requirements: whether terrains accessible to current and
near-future missions retain radiolytic donor and acceptor products long
enough for that supply to be biologically useful, and whether the
resulting energy flux is large enough to sustain detectable cell
abundance (Atri et al., 2025; Hoehler and Jørgensen, 2013; Bradley et
al., 2020). Both requirements can be
addressed using measurement pathways demonstrated or proposed for in
situ and returned-sample analyses (Sutter et al., 2017; Goesmann et al.,
2017; Hou et al., 2025).

Three terrain-dependent variables govern the modeled energy supply: the
fraction of water-equivalent hydrogen (WEH) available to radiolysis, the
retention and loss of radiolytically produced H₂, and the abundance and
accessibility of electron acceptors (Fe(III), sulfate, nitrate,
oxychlorines, and direct radiolytic oxidants) (Mitrofanov et al., 2022;
Lasne et al., 2016; Dzaugis et al., 2018). In the parameter space explored
here, H₂ retention and loss produce the largest change in
protected-depth radiolytic redox power and determine the active
or inactive state of a terrain, consistent with the central role of H₂
production and retention in radiolytic habitability frameworks (Dzaugis
et al., 2018; Tarnas et al., 2021; Atri et al., 2025). We then convert
the available redox power into a maximum supported microbial cell
density, using the maintenance-power framework commonly applied in
deep-biosphere energetics (Hoehler and Jørgensen, 2013; LaRowe and
Amend, 2015; Bradley et al., 2020). This conversion places the energy supply and biological maintenance
demand in comparable units.

The terrains considered here are motivated by current and near-future
mission planning (Quantin-Nataf et al., 2021; Hou et al., 2025; Vago et
al., 2017). Oxia Planum is the landing site for ESA's ExoMars Rosalind Franklin
rover, which carries a 2 m drill into clay-bearing Noachian stratigraphy
and the MOMA pyrolysis instrument (Quantin-Nataf et al., 2021; Goesmann
et al., 2017; Vago et al., 2017). Southern Utopia is one of the Tianwen-3
sample-return reference terrains (Hou et al., 2025). Arabia/Mawrth is
included as a clay-rich representative case for strongly hydrated
terrains in the broader Mars Sample Return target window, following
orbital evidence for extensive phyllosilicate-bearing Noachian terrains
in Mawrth Vallis and Arabia-adjacent regions (Poulet et al., 2005;
Bishop et al., 2008; Quantin-Nataf et al., 2021). Gale Crater, by
contrast, has been characterized in detail by Curiosity over more than six
martian years, including direct evolved-gas analyses of drilled
subsurface samples by SAM (Hassler et al., 2014; Mitrofanov et al.,
2022; Sutter et al., 2017), and provides a Curiosity/DAN-constrained
reference case for evaluating shallow-radiolysis models.

The present study provides the integration and the biological conversion
by combining shallow GCR transport, radiolysis chemistry, H₂ retention,
and maintenance-power scaling in the RHZ context (Atri et al., 2025;
Hoehler and Jørgensen, 2013; Bradley et al., 2020). We report the water
fraction available to radiolysis, the retained donor and oxidant
abundances, the protected-depth redox power, and the
maintenance-power-derived maximum supported cell density for these four
terrains. We also identify H₂ retention and loss as the principal
physical sensitivity and specify a measurement strategy that can
distinguish the modeled active and inactive states in relevant mission
or returned-sample contexts.

\section{2. Methods}\label{methods}

The model couples four physical components: Monte Carlo transport of the
galactic cosmic-ray (GCR) cascade through the martian atmosphere and
regolith (Section 2.1); phase-resolved post-radiolysis chemistry that
converts energy deposition into retained reduced and oxidized species
(Section 2.2); a depth-resolved evaluator that expresses the retained
chemistry as a catabolic free-energy flux, the radiolytic redox power
(Section 2.3); and a conversion of that flux to a maximum supported cell
density through a per-cell maintenance power (Section 2.4). The transport
stage uses the Geant4 toolkit (Agostinelli et al., 2003; Allison et al.,
2006; Allison et al., 2016); the chemistry, energetics, and cell-density
stages are deterministic. The governing equations, source terms, and
parameter values for each component are specified in full below, so that
the model can be reproduced from the physics and chemistry given here.

\subsection{2.1 Radiation Transport and Water Available to
Radiolysis}\label{radiation-transport-and-water-available-to-radiolysis}

GCR transport through 2 m martian regolith columns at 1 cm depth
resolution is simulated with Geant4 (Agostinelli et al., 2003; Allison
et al., 2006; Allison et al., 2016). The physics configuration combines
the QGSP\_BERT\_HP hadronic list (a quark-gluon string model at high
energy with a Bertini intranuclear cascade at intermediate energy and
high-precision transport of thermal and epithermal neutrons) with the
high-accuracy electromagnetic option (Geant4 EM option 4) and the
standard ion and radioactive-decay processes. The high-accuracy
electromagnetic option is used because low-energy electron and delta-ray
transport sets the ionization density that controls radiolytic yields.
Geant4 tracks each primary and its full secondary cascade and records
energy deposition, particle fluence, and ionization yield by particle
species in each 1 cm layer. The geometry comprises a 20-layer
exponentially stratified Mars atmosphere (CO₂-dominant, scale height
11.1 km, surface pressure ≈ 600 Pa) overlying a 200-layer homogeneous or
two-layer regolith column. Scenario compositions specify oxide
chemistry, density, total water-equivalent hydrogen, neutron absorbers,
and trace capture elements.

Primary particles are sampled from the four dominant GCR components:
protons (87\%), helium (12\%), carbon (0.5\%), and iron (0.5\%) by
particle abundance, following the relative abundances of Usoskin et
al.~(2005). The differential energy spectrum of each species is the
local interstellar spectrum modulated to the inner heliosphere by the
Gleeson and Axford (1968) force-field approximation, in which the
modulated kinetic energy is shifted from the interstellar value by
\(T_{\mathrm{LIS}} = T + (Z/A)\,\phi\), with a single modulation
potential φ = 660 MV representative of the Curiosity/RAD comparison epoch
(Gleeson and Axford, 1968; Usoskin et al., 2005; Hassler et al., 2014;
Guo et al., 2015). Primaries are launched over the downward hemisphere
with a cosine (Lambert) angular weighting, which reproduces an isotropic
GCR field as a one-way hemispheric flux \(J_{\mathrm{GCR}}\). Solar-cycle
variation enters only through φ and is treated as an external
multiplicative transport-context uncertainty; elevation-scaled values of
\(J_{\mathrm{GCR}}\) for each terrain are given in Section 3.1.

A thin silicon slab above the regolith provides a RAD-like
daily-averaged surface-dose normalization check, and upward neutron
leakage is recorded as a DAN-like moderation diagnostic; both serve as
transport-level comparisons against in situ landed measurements rather
than as habitability metrics.

Total WEH is distinguished from the water fraction available to
radiolysis. Structural OH and tightly bound water are excluded from the
radiolytically active water pool, because the biologically relevant
radiolytic chemistry occurs in liquid-like or transient hydrated phases
rather than in immobile lattice sites. Hydrated cases are interpreted as
interlayer-derived transient hydrated-film states, not as biology
occurring in structural or immobile interlayer water. Brine cases
represent transient liquid-water states with reduced water activity and
shorter residence. The ClO₄⁻-equivalent term is a reactive oxychlorine
proxy rather than a direct measured perchlorate abundance.

\subsection{2.2 Post-Radiolysis
Chemistry}\label{post-radiolysis-chemistry}

For each 1 cm layer, the radiolytic production rate of species \(i\) is
obtained by weighting the energy deposited by each particle class \(j\)
by a phase- and LET-dependent radiolytic yield (G value),
\[
S_i = \sum_j G_{i,j}\,\dot{E}_j ,
\]
where \(\dot{E}_j\) is the annual energy deposition from class \(j\) in
the radiolytically active water (or oxychlorine) pool and \(G_{i,j}\) is
expressed in molecules per 100 eV. Water-radiolysis yields follow the
liquid-water values of LaVerne (2000) and the molecular-hydrogen/peroxide
coupling of Pastina and LaVerne (2001): G(H₂) increases with linear
energy transfer (LET) from 0.45 for protons, electrons, and γ through
1.10 for α particles to 1.50 for heavy ions, whereas G(H₂O₂) is
0.70 for electrons and γ, 0.87--0.98 for neutrons and protons, and falls
at high LET to 0.38 (α) and 0.15 (heavy ions). Water-ice
yields (enhanced G(H₂); suppressed high-LET G(H₂O₂)) are substituted for
frost cases, and brine is treated as liquid water to within ±15\%.
Oxychlorine radiolysis is represented as an effective ClO₄⁻-equivalent
source with G(O₂) ≈ 0.012--0.035 and G(ClO₃⁻) ≈ 0.008--0.028 molecules
per 100 eV across the LET range, consistent with Mars-analog perchlorate
irradiation (Quinn et al., 2013) but carrying a factor-of-2--3 systematic
uncertainty that is propagated explicitly (Section 2.6).

The Tier 1 post-radiolysis chemistry is a deterministic per-layer box
model. Molecular hydrogen partitions between an aqueous/sorbed reservoir
and a gas reservoir and is lost from the gas phase by first-order escape,
\[
\frac{dC_{\mathrm{H_2}}^{\mathrm{aq}}}{dt} = S_{\mathrm{H_2}}
  - k_{\mathrm{p}}C_{\mathrm{H_2}}^{\mathrm{aq}}
  + k_{\mathrm{r}}C_{\mathrm{H_2}}^{\mathrm{gas}}, \qquad
\frac{dC_{\mathrm{H_2}}^{\mathrm{gas}}}{dt} =
    k_{\mathrm{p}}C_{\mathrm{H_2}}^{\mathrm{aq}}
  - k_{\mathrm{r}}C_{\mathrm{H_2}}^{\mathrm{gas}}
  - k_{\mathrm{esc}}C_{\mathrm{H_2}}^{\mathrm{gas}},
\]
with reference escape rate \(k_{\mathrm{esc}} = 0.20\ \mathrm{yr^{-1}}\)
(Section 2.5). Hydrogen peroxide is removed by an effective Fe-mediated
(Fenton) sink and by surface-catalyzed disproportionation,
\[
\frac{dC_{\mathrm{H_2O_2}}}{dt} = S_{\mathrm{H_2O_2}}
  - (k_{\mathrm{Fenton}} + k_{\mathrm{surf}})\,C_{\mathrm{H_2O_2}},
\]
with the disproportionation branch returning oxygen at a rate
\(\tfrac{1}{2}\,k_{\mathrm{surf}}C_{\mathrm{H_2O_2}}\). Oxygen partitions
and escapes by the same two-reservoir law as H₂. Oxychlorine inventory
evolves through the dominant radiolytic branch
ClO₄⁻ → ClO₃⁻ + ½ O₂, with chlorate and O₂ produced at their respective
radiolytic G-value rates and the ClO₄⁻ reservoir depleted by chlorine
mass balance; chlorate carries no default loss term and persists as an
oxidant reservoir. The thermodynamically favorable donor--acceptor
reactions (H₂ + H₂O₂, H₂ + ½ O₂, and the H₂/oxychlorine reactions) are
not imposed as abiotic kinetic sinks, because at cold martian regolith
temperatures they are not expected to proceed without catalysis or
biology; they enter only as the energy-bookkeeping pairings of the
redox-power evaluator (Section 2.3).

A Tier 2 configuration adds the short-lived aqueous radical
intermediates (the H, OH, HO₂, and superoxide radicals and the hydrated
electron) as mobile spur-escape fractions; it is validated against
radical limiting cases and used as a robustness check, given that
Mars-relevant radical kinetics remain underconstrained at the ppb-level
intermediates active here. The deterministic chemistry solver passes all
fourteen conservation, stoichiometric, and thermodynamic consistency
checks, including H₂ mass balance, analytic–numerical agreement for
peroxide, Tier 2 collapse to Tier 1 at zero radicals, and the expected
reaction-energetics sign tests.

\subsection{2.3 Radiolytic Redox Power}\label{radiolytic-redox-power}

Depth-resolved radiolytic redox power is introduced here to connect
radiation physics and microbial energetics. Previous RHZ studies by Atri
(2016, 2020) and Atri et al.~(2025), together with related Mars
radiolysis and radiation-environment studies, have reported GCR
ionization, dose, electron-production, or H₂-production terms. These
quantities capture important radiolytic physics, but they are not
directly comparable to biological maintenance-power requirements:
production terms do not account for geochemical partitioning and
thermodynamic availability, while absorbed doses are not in the currency
of microbial metabolism. A meaningful comparison requires evaluating, at
each depth and for each candidate catabolic reaction, the Gibbs free
energy available under Mars in situ conditions, weighted by the
electron-transfer capacity that the retained chemistry can sustain. The redox-power calculation provides this conversion by pairing
retained donor and acceptor abundances at each depth, weighting the
matched electron-transfer capacity by the in situ \(\Delta G\) of the
corresponding catabolic half-reaction (computed as
\(\Delta G = \Delta G^\circ + RT \ln Q\) under Mars-conditioned
temperature, water activity, pH, and representative dissolved-species
activities), and normalizing to the bulk regolith mass. The result is a
depth-resolved catabolic power density in µJ kg⁻¹ yr⁻¹ of bulk regolith
that can be divided directly by a per-cell maintenance power to yield a
maximum supported cell density. Normalization to bulk regolith mass also
ensures inter-terrain comparability across sites differing in density,
water content, and acceptor mineralogy.

Operationally, the calculation reports the chemically useful annual
energy at depth \(z\) in µJ kg⁻¹ yr⁻¹ of bulk regolith, where the
normalization mass includes solids and the modeled water content. For
each depth and reaction family, the evaluator pairs H₂ donor
availability with the corresponding H₂-equivalent acceptor capacity,
multiplies by the Mars-conditioned available energy per pairing (the
in situ Gibbs energy \(\Delta G = \Delta G^\circ + RT \ln Q\) evaluated
at the scenario temperature, water activity, pH, ionic strength, and
representative dissolved-species activities), and applies
water-activity, residence/accessibility, and organic-preservation
factors. Reaction families considered are H₂/H₂O₂, H₂/O₂, H₂/ClO₃⁻,
H₂/ClO₄⁻, H₂/Fe(III), H₂/SO₄²⁻, and H₂/NO₃⁻.

The reference calculation reports annual source-limited redox power
conditional on sorbed or mineral-associated H₂ retention. A second,
one-year retained-inventory formulation uses the retained H₂ abundance
as the donor availability and supports H₂-loss sensitivity tests.
Main-text redox-power values use a protected-depth selector of ≥10 cm, a
postprocessing policy motivated by the dose-depth profiles of Pavlov et
al., the oxidant-depth constraints of Lasne et al., and ultraviolet
attenuation, while remaining within the access depth of the ExoMars Rosalind
Franklin drill and standard returned-sample protocols.

\subsection{2.4 Maintenance-Power Conversion to Supported Cell
Density}\label{maintenance-power-conversion-to-supported-cell-density}

Cellular viability is set by whether the catabolic free-energy flux a
cell can access exceeds the minimum power it must dissipate to remain
alive, the maintenance or basal power requirement. The redox power
computed above is normalized per kilogram of rock, so two of its
properties must be fixed before it can be compared to that per-cell
requirement.

First, the reported redox power is already a catabolic free-energy flux,
not a deposited-dose flux. The conversion from absorbed GCR energy to
retained radiolytic product is carried by the G values in the chemistry
solver (Section 2.2), and the evaluator then multiplies each retained
donor-acceptor pairing by the Mars-conditioned Gibbs energy of the
corresponding catabolic reaction (Section 2.3). The reported value is
therefore the in situ \(\Delta G\) flux available from H₂ oxidation, the
same quantity in which microbial maintenance power is defined. No
additional biological-efficiency factor is applied between redox power
and maintenance power, because maintenance power is itself the catabolic
power a cell must consume to persist and already embeds the inefficiency
of energy conservation. For the active cases, the dominant pairing is
\(\mathrm{H_2 + H_2O_2 \rightarrow 2H_2O}\), with in situ \(\Delta G\)
of approximately \(-310\ \mathrm{kJ\ mol^{-1}}\) of H₂ at default Mars
shallow-subsurface conditions (210 K, \(a_w = 0.85\),
\(a_{\mathrm{H_2}} = 10^{-6}\), \(a_{\mathrm{H_2O_2}} = 10^{-6}\); the
standard-state value is \(-340\ \mathrm{kJ\ mol^{-1}}\)). This is larger
than both the biological energy quantum (about
\(-20\ \mathrm{kJ\ mol^{-1}}\)) and plausible corrections to
dissolved-species activities, so the available energy per reaction is
not expected to be order-of-magnitude sensitive to those concentration
assumptions.

Second, the conversion is bulk-averaged, and the two quantities that
carry physical meaning are invariant to where the chemistry and the
cells reside. The supported cell-density estimate follows the
deep-biosphere energetic formulation of Hoehler and Jørgensen (2013) and
LaRowe and Amend (2015):

\[
N_{\max} = \frac{P_{\mathrm{redox}}}{P_{\mathrm{maint}}}\rho_{\mathrm{bulk}}
\]

where \(P_{\mathrm{maint}}\) is the per-cell maintenance power,
\(P_{\mathrm{redox}}\) is the redox power expressed as W kg⁻¹ of bulk
regolith, \(\rho_{\mathrm{bulk}}\) is the terrain-specific bulk density
in kg cm⁻³, and \(N_{\max}\) is the maximum sustainable cell density in
cells cm⁻³ of bulk regolith. Redox power in µJ kg⁻¹ yr⁻¹ is converted to
W kg⁻¹ by dividing by \(3.156 \times 10^7\ \mathrm{s\ yr^{-1}}\), and
cell densities are reported using terrain-specific bulk densities (2.20,
2.25, and 1.70 g cm⁻³ for Oxia, Arabia/Mawrth, and southern Utopia,
respectively).
The per-cell power that determines survival is the total energy produced
divided by the total cells supported, and the total supported cell
abundance per unit terrain is the same energy divided by
\(P_{\mathrm{maint}}\); both are invariant to whether the energy and the
cells are treated as bulk-distributed or as confined to the thin
hydrated film, because the film mass appears in the numerator (energy
produced) and is tracked unchanged to the denominator (cells supported).
The only quantity that depends on that choice is the local cell
concentration within the habitable film, which exceeds the bulk-averaged
value by the inverse film volume fraction. For the volume fractions of
water available to radiolysis in the three active terrains, 8.5\% for
Oxia (WEH 5.5 wt\%, fraction available to radiolysis 0.70, bulk density
2.20 g cm⁻³), 7.3\% for Arabia/Mawrth (WEH 6.5 wt\%, 0.50, 2.25 g cm⁻³),
and 3.6\% for southern Utopia (WEH 7.0 wt\%, 0.30, 1.70 g cm⁻³), this
factor is approximately 12, 14, and 28 respectively. We report the
bulk-averaged density throughout, because it is the quantity directly
comparable to bulk-normalized terrestrial deep-biosphere cell counts and
because survivability is set by per-cell power, not by local packing.

The result is a steady-state maximum supported cell density: it assumes
every joule of catabolic free energy is available for maintenance, with
nothing lost to growth, cell decay, or abiotic sinks. Because
empirically inferred and modeled maintenance powers span roughly five
orders of magnitude, \(P_{\mathrm{maint}}\) is varied across a range rather than
fixed, at \(10^{-15}\) W (culture-based laboratory maintenance),
\(10^{-18}\) W (the oxic upper end of subseafloor cell-specific power),
and \(10^{-20}\) W (the anaerobic subseafloor median and survival
floor), following Tijhuis et al.~(1993), LaRowe and Amend (2015), and
Bradley et al.~(2020). Radiolytic redox power remains the primary
metric; \(N_{\max}\) is reported as a derived biological constraint, not
a second index.

\subsection{2.5 Hydrogen Escape Prior and Sensitivity
Cases}\label{hydrogen-escape-prior-and-sensitivity-cases}

The reference H₂ escape coefficient, \(k = 0.20\ \mathrm{yr^{-1}}\), is
a plausibility-based effective first-order loss rate, not a direct
measurement. A first-mode slab estimate,

\[
k = \frac{\pi^2 D_{\mathrm{eff}}}{L^2},
\]

with \(L = 10\ \mathrm{cm}\) gives
\(D_{\mathrm{eff}} = 6.4 \times 10^{-12}\ \mathrm{m^2\ s^{-1}}\) at
\(k = 0.20\ \mathrm{yr^{-1}}\). This diffusivity lies far below the open
gas-pore H₂ diffusion coefficient and represents a sorbed,
mineral-associated, tortuous, or otherwise retarded H₂-retention state.
The Latin-hypercube prior samples a multiplicative escape-rate factor
from 0.1 to 10, equivalent to \(k = 0.02\) to \(2.0\ \mathrm{yr^{-1}}\).
Because H₂ retention and loss constitute the largest sampled sensitivity
for shallow radiolytic activity (Section 3.4), they are examined under
two sensitivity configurations: gas-phase escape variation within
baseline aqueous/sorbed partitioning, and a connected-pore free-gas case
in which newly produced H₂ is routed directly into the gas reservoir
without redissolution.

\subsection{2.6 Uncertainty Propagation}\label{uncertainty-propagation}

Uncertainty propagation uses Latin-hypercube sampling over cached 10
000-event transport and chemistry outputs. Sampled controls are
free-water fraction, oxychlorine fraction, water G-value scale,
oxychlorine G-value scale, peroxide sink scale, H₂ escape scale, brine
residence fraction, pH, and water activity. Protected-depth summaries
use 1024 samples per case. Two independent 1024-sample sets, with a
pooled 2048-sample diagnostic, give active-case median differences of
0.2--8.1\% and consistently identify the H₂ escape factor as the largest
sampled control.

\section{3. Results}\label{results}

\subsection{3.1 Radiolytic Donor and Oxidant Abundances Across
Terrains}\label{radiolytic-donor-and-oxidant-abundances-across-terrains}

Across the four reference terrains, the radiolytic donor abundance is
set primarily by the water fraction available to radiolysis, while the
direct oxidant abundance is set by both water available to radiolysis
(for H₂O₂ and O₂) and the oxychlorine pool (for chlorate). The seven
canonical 10 000-particle cases (Gale/Bradbury composite, Oxia hydrated,
Oxia brine, southern Utopia hydrated, southern Utopia brine,
Arabia/Mawrth hydrated, and Arabia/Mawrth brine) span more than two
orders of magnitude in H₂ source rate and approach two orders of
magnitude in retained donor/acceptor balance (Table 1). Gale uses
\(J_{\mathrm{GCR}} = 2.30\ \mathrm{cm^{-2}\ s^{-1}}\) and a 5 cm
transient-water cap; Oxia, southern Utopia, and Arabia/Mawrth use
elevation-scaled fluxes of 2.50, 2.38, and 2.52 cm⁻² s⁻¹. Surface
silicon benchmark ratios span 0.939--1.039 relative to the adopted
RAD-like reference, confirming that inter-terrain differences in
radiolytic output are driven by partition and chemistry, not by
transport-level disagreement with landed dosimetry.

\textbf{Table 1. One-year radiolysis products by terrain.} Retained
oxidant H₂-equivalent is computed from direct retained radiolytic
oxidants as H₂O₂ + 2 O₂ + 3 ClO₃⁻. Direct donor/acceptor ratios are
retained H₂ divided by this stoichiometric H₂-equivalent capacity and
exclude geological acceptor reservoirs.

{\footnotesize\setlength{\tabcolsep}{3pt}\def\LTcaptype{none} 
\begin{longtable}[]{@{}
  >{\raggedright\arraybackslash}p{(\linewidth - 10\tabcolsep) * \real{0.1364}}
  >{\raggedright\arraybackslash}p{(\linewidth - 10\tabcolsep) * \real{0.1364}}
  >{\raggedleft\arraybackslash}p{(\linewidth - 10\tabcolsep) * \real{0.1818}}
  >{\raggedleft\arraybackslash}p{(\linewidth - 10\tabcolsep) * \real{0.1818}}
  >{\raggedleft\arraybackslash}p{(\linewidth - 10\tabcolsep) * \real{0.1818}}
  >{\raggedleft\arraybackslash}p{(\linewidth - 10\tabcolsep) * \real{0.1818}}@{}}
\toprule\noalign{}
\begin{minipage}[b]{\linewidth}\raggedright
Case
\end{minipage} & \begin{minipage}[b]{\linewidth}\raggedright
Phase
\end{minipage} & \begin{minipage}[b]{\linewidth}\raggedleft
H₂ source (nmol m⁻² yr⁻¹)
\end{minipage} & \begin{minipage}[b]{\linewidth}\raggedleft
Retained H₂ (nmol m⁻²)
\end{minipage} & \begin{minipage}[b]{\linewidth}\raggedleft
Retained oxidant H₂-eq (nmol m⁻²)
\end{minipage} & \begin{minipage}[b]{\linewidth}\raggedleft
Direct donor/acceptor ratio
\end{minipage} \\
\midrule\noalign{}
\endhead
\bottomrule\noalign{}
\endlastfoot
Gale/Bradbury & composite & 1.45 & 1.41 & 5.99 & 0.235 \\
Oxia & hydrated film & 313 & 307 & 6.96 & 44.1 \\
Oxia & brine & 313 & 301 & 10.6 & 28.3 \\
Southern Utopia & hydrated film & 157 & 153 & 6.26 & 24.5 \\
Southern Utopia & brine & 157 & 151 & 9.50 & 15.9 \\
Arabia/Mawrth & hydrated film & 270 & 265 & 7.17 & 36.9 \\
Arabia/Mawrth & brine & 270 & 260 & 11.0 & 23.6 \\
\end{longtable}
}

Once geological acceptor reservoirs (Fe(III), sulfate, nitrate, and the
oxychlorine proxy) are included in the evaluator, all active cases
become donor limited rather than acceptor limited. The dominant
operative reaction across every active case is H₂ + H₂O₂ (Table 2).

\subsection{3.2 Gale: a Curiosity/DAN-Constrained Null at Protected
Depth}\label{gale-a-curiositydan-constrained-null-at-protected-depth}

The Gale/Bradbury composite is radiolytically inactive at protected
depth (≥10 cm) in the present-day configuration (Table 2, top row). The
mechanism of the null is direct. Under the partition of water available
to radiolysis consistent with Curiosity/DAN constraints, radiolysable
transient water is confined to the upper 5 cm, so the protected-depth
selector contains no active radiolysable-water layer. The raw column H₂
source rate is approximately 1.45 nmol m⁻² yr⁻¹, two orders of magnitude
below the active terrains, and the resulting one-year retained donor
abundance is 1.41 nmol m⁻² (Table 1), but this inventory is not available
at the protected depths used for the activity test. Once water activity
and residence/accessibility are applied, none of the seven candidate
reaction families exceeds the available-energy threshold at protected
depth. This is a model result for a
single, well-instrumented terrain. It applies to a Gale-like
water-availability and oxychlorine partition, which does not meet the
modeled activity threshold, and not to the shallow martian subsurface in
general. As the most directly
Curiosity/DAN-constrained anchor in this study, the Gale case provides
a conservative reference against which the active terrains are evaluated
and constrains the interpretation of the active predictions as
terrain-dependent rather than generic to the martian subsurface.

The Gale null is independently consistent with published in situ
evolved-gas analyses. SAM stepped-temperature pyrolysis on multiple Gale
samples (Rocknest aeolian fines, the John Klein and Cumberland
mudstones, and several Murray- and Stimson-formation drill targets) has
detected H₂ release across a wide temperature range, but the
high-temperature H₂ peaks above approximately 400 °C are quantitatively
consistent with structural H release from Fe²⁺-bearing phyllosilicates
by oxidative dehydrogenation rather than with radiolytically retained H₂
(Sutter et al., 2017; Lempart et al., 2020). Lower-temperature H₂ in
several samples is attributed by Sutter et al.~(2017) to water
fragmentation in the SAM mass spectrometer rather than to a radiolytic
source. No published SAM dataset to date exhibits a
low-to-intermediate-temperature H₂ excess above this background that
would require a radiolytically retained reservoir at the levels
associated with our active terrains. The current SAM record at Gale is
therefore consistent with the protected-depth null reported here.

\subsection{3.3 Active Terrains: Oxia Planum, Arabia/Mawrth, and
Southern
Utopia}\label{active-terrains-oxia-planum-arabiamawrth-and-southern-utopia}

The three active terrains all sustain non-zero protected-depth
radiolytic redox power under shared sorbed-H₂ priors (Table 2). Under
the reference donor formulation, hydrated-film cases yield median
protected redox powers of 19.1 µJ kg⁻¹ yr⁻¹ of bulk regolith (Oxia),
17.2 µJ kg⁻¹ yr⁻¹ (Arabia/Mawrth), and 11.9 µJ kg⁻¹ yr⁻¹ (southern
Utopia). The dominant operative reaction in every active case is H₂ +
H₂O₂, and the limiting factor in every active case is donor supply.

\textbf{Table 2. Protected-depth radiolytic redox power by terrain.}
Values use the ≥10 cm protected-depth selector and reservoir-inclusive
matching.

{\footnotesize\setlength{\tabcolsep}{3pt}\def\LTcaptype{none} 
\begin{longtable}[]{@{}
  >{\raggedright\arraybackslash}p{(\linewidth - 10\tabcolsep) * \real{0.1500}}
  >{\raggedleft\arraybackslash}p{(\linewidth - 10\tabcolsep) * \real{0.2000}}
  >{\raggedleft\arraybackslash}p{(\linewidth - 10\tabcolsep) * \real{0.2000}}
  >{\raggedright\arraybackslash}p{(\linewidth - 10\tabcolsep) * \real{0.1500}}
  >{\raggedright\arraybackslash}p{(\linewidth - 10\tabcolsep) * \real{0.1500}}
  >{\raggedright\arraybackslash}p{(\linewidth - 10\tabcolsep) * \real{0.1500}}@{}}
\toprule\noalign{}
\begin{minipage}[b]{\linewidth}\raggedright
Case
\end{minipage} & \begin{minipage}[b]{\linewidth}\raggedleft
Protected depth (cm)
\end{minipage} & \begin{minipage}[b]{\linewidth}\raggedleft
Median redox power (µJ kg⁻¹ yr⁻¹ bulk regolith)
\end{minipage} & \begin{minipage}[b]{\linewidth}\raggedright
5--95\% interval
\end{minipage} & \begin{minipage}[b]{\linewidth}\raggedright
Dominant reaction
\end{minipage} & \begin{minipage}[b]{\linewidth}\raggedright
Limiting factor
\end{minipage} \\
\midrule\noalign{}
\endhead
\bottomrule\noalign{}
\endlastfoot
Gale/Bradbury composite & 10 & 0 & 0--0 & none & none \\
Oxia hydrated film & 13 & 19.1 & 2.30--160 & H₂ + H₂O₂ & donor \\
Arabia/Mawrth hydrated film & 10 & 17.2 & 2.01--145 & H₂ + H₂O₂ &
donor \\
Southern Utopia hydrated film & 12 & 11.9 & 1.34--99.3 & H₂ + H₂O₂ &
donor \\
Oxia brine & 13 & 1.53 & 0.107--16.3 & H₂ + H₂O₂ & donor \\
Arabia/Mawrth brine & 10 & 1.35 & 0.112--17.0 & H₂ + H₂O₂ & donor \\
Southern Utopia brine & 12 & 0.949 & 0.0730--10.3 & H₂ + H₂O₂ & donor \\
\end{longtable}
}

The independent 5--95\% intervals for the three hydrated cases overlap
strongly. A paired shared-prior diagnostic returns
\(P(\mathrm{Oxia} > \mathrm{Arabia/Mawrth}) = 1.00\),
\(P(\mathrm{Oxia} > \mathrm{southern\ Utopia}) = 1.00\), and
\(P(\mathrm{Arabia/Mawrth} > \mathrm{southern\ Utopia}) = 1.00\),
because the largest sampled uncertainty acts as a common multiplicative
H₂-escape control. The paired diagnostic does not include independent
terrain-specific systematic errors in the accessibility of water
available to radiolysis, sorption state, or acceptor mineralogy. Within
the paired shared-prior experiment, the ordering of median redox power
is preserved; however, the absolute uncertainty envelope is too broad
for this metric alone to determine landing-site or sample-prioritization
decisions. We return to this limit in Section 4.5.

\subsection{3.4 H₂ Retention and Loss Control the Active-State
Transition}\label{hux2082-retention-and-loss-control-the-active-state-transition}

H₂ retention and loss constitute the modeled physical factor that
produces the largest change in protected redox power across the active
terrains, and they are the main distinction between active and inactive
model states.

Within the baseline aqueous/sorbed partitioning, increasing the
gas-phase escape rate from \(k = 0.20\ \mathrm{yr^{-1}}\) to
\(365\ \mathrm{yr^{-1}}\) and \(8760\ \mathrm{yr^{-1}}\) reduces
hydrated retained-inventory redox power from 11.03, 6.28, and 9.49 µJ
kg⁻¹ yr⁻¹ to 8.08, 4.60, and 6.96, and then to 8.07, 4.59, and 6.94, for
Oxia, southern Utopia, and Arabia/Mawrth respectively. The signal is
degraded but not eliminated, because newly produced H₂ remains buffered
in the aqueous/sorbed reservoir before escaping.

The connected-pore free-gas case behaves qualitatively differently. With
newly produced H₂ routed directly into the gas reservoir and
redissolution disabled, hydrated retained-inventory protected redox
power decreases to 0.0387, 0.0220, and 0.0333 µJ kg⁻¹ yr⁻¹ at
\(k = 365\ \mathrm{yr^{-1}}\), and to 0.00236, 0.00134, and 0.00203 at
\(k = 8760\ \mathrm{yr^{-1}}\), for Oxia, southern Utopia, and
Arabia/Mawrth. This is a reduction of approximately \(285\) to
\(4700\times\) relative to the sorbed baseline at
\(k = 0.20\ \mathrm{yr^{-1}}\). The retained/source ratios in this case
(\(2.74 \times 10^{-3}\) at \(k = 365\ \mathrm{yr^{-1}}\) and
\(1.14 \times 10^{-4}\) at \(k = 8760\ \mathrm{yr^{-1}}\)) match
first-order gas-loss expectations, confirming the reduction is
mechanistic rather than numerical.

The resulting model contrast is primarily physical. Active predictions
for shallow radiolytic habitability require sorbed or mineral-associated
H₂ retention. In the connected-pore free-gas case, every active terrain
falls below the protected-depth activity threshold. The active or
inactive model state is therefore more sensitive to H₂ partitioning than
to the total radiolytic H₂ source, which differs by less than a factor
of three across the active terrains. The decisive variable is
therefore a measurable property of the host material: the partitioning
of free pore space against sorption-active mineral surface area.

\subsection{3.5 Maintenance-Power Conversion to Supported Cell
Density}\label{maintenance-power-conversion-to-supported-cell-density-1}

The redox power above is only biologically interpretable once it is
converted to a maintenance-power constraint. Expressed as power density,
the active hydrated terrains deliver 6.1 × 10⁻¹³ W kg⁻¹ of bulk regolith
(Oxia), 5.5 × 10⁻¹³ (Arabia/Mawrth), and 3.8 × 10⁻¹³ (southern Utopia).
Dividing by per-cell maintenance power gives the maximum supported cell
density (Table 3).

\textbf{Table 3. Maximum supported cell density for the active hydrated
terrains.} Values are steady-state energetic maxima,
\(N_{\max} = (P_{\mathrm{redox}}/P_{\mathrm{maint}})\rho_{\mathrm{bulk}}\),
reported as cells cm⁻³ of bulk regolith at terrain-specific bulk
densities (2.20, 2.25, and 1.70 g cm⁻³ for Oxia, Arabia/Mawrth, and
southern Utopia, respectively).
\(P_{\mathrm{maint}}\) spans culture-based laboratory maintenance
(\(10^{-15}\) W), the oxic upper end of empirically inferred subseafloor
cell-specific power (\(10^{-18}\) W), and the anaerobic subseafloor
median and survival floor (\(10^{-20}\) W; Bradley et al., 2020).

{\footnotesize\setlength{\tabcolsep}{3pt}\def\LTcaptype{none} 
\begin{longtable}[]{@{}
  >{\raggedright\arraybackslash}p{(\linewidth - 8\tabcolsep) * \real{0.1579}}
  >{\raggedleft\arraybackslash}p{(\linewidth - 8\tabcolsep) * \real{0.2105}}
  >{\raggedleft\arraybackslash}p{(\linewidth - 8\tabcolsep) * \real{0.2105}}
  >{\raggedleft\arraybackslash}p{(\linewidth - 8\tabcolsep) * \real{0.2105}}
  >{\raggedleft\arraybackslash}p{(\linewidth - 8\tabcolsep) * \real{0.2105}}@{}}
\toprule\noalign{}
\begin{minipage}[b]{\linewidth}\raggedright
Terrain (hydrated)
\end{minipage} & \begin{minipage}[b]{\linewidth}\raggedleft
Power density (W kg⁻¹ bulk regolith)
\end{minipage} & \begin{minipage}[b]{\linewidth}\raggedleft
\(N_{\max}\) at \(10^{-15}\) W (cm⁻³)
\end{minipage} & \begin{minipage}[b]{\linewidth}\raggedleft
\(N_{\max}\) at \(10^{-18}\) W (cm⁻³)
\end{minipage} & \begin{minipage}[b]{\linewidth}\raggedleft
\(N_{\max}\) at \(10^{-20}\) W (cm⁻³)
\end{minipage} \\
\midrule\noalign{}
\endhead
\bottomrule\noalign{}
\endlastfoot
Oxia & 6.1 × 10⁻¹³ & ≈1 & ≈1 × 10³ & ≈1 × 10⁵ \\
Arabia/Mawrth & 5.5 × 10⁻¹³ & ≈1 & ≈1 × 10³ & ≈1 × 10⁵ \\
Southern Utopia & 3.8 × 10⁻¹³ & ≈0.6 & ≈6 × 10² & ≈6 × 10⁴ \\
\end{longtable}
}

The biological implication is a low maximum supported cell density. At
culture-based laboratory maintenance powers, the retained-H₂ RHZ
supports of order one cell per cm³ or fewer. The active terrains reach
measurable supported densities only if putative inhabitants operate at
or near the lower end of empirically inferred subseafloor power
requirements, at which point the energetic maximum rises to order 10³
cells cm⁻³ at the oxic upper end and order 10⁵ cells cm⁻³ at the
anaerobic survival floor. The brine cases, an order of magnitude lower
in redox power, shift the entire estimate downward by a further factor
of ten.

This places the modeled RHZ in a specific and independently
characterized biological setting, and the two parts of that comparison
should be kept separate. The power level is anchored by empirical
estimates from Earth's energy-limited subsurface. Cell-specific power in
global subseafloor sediments has a median of \(3.3 \times 10^{-20}\) W,
with sulfate reducers near \(1 \times 10^{-19}\) W and aerobic
heterotrophs near \(2 \times 10^{-18}\) W (Bradley et al., 2020), so the
survival end of the \(P_{\mathrm{maint}}\) range is observed rather
than assumed. At the sulfate-reducer power, the metabolism most directly
analogous to the H₂-driven RHZ chemistry, the active terrains support of
order 10⁴ cells cm⁻³ (\(6 \times 10^3\) to \(1 \times 10^4\) across the
three terrains). The relevant terrestrial analog is mechanistic rather
than quantitative. The radiolytic-hydrogen-fed crustal biome of the
Witwatersrand Basin, dominated by the self-sufficient sulfate reducer
\textit{Candidatus Desulforudis audaxviator}, subsists on H₂ produced by
radiolysis (Lin et al., 2006; Chivian et al., 2008). That biome
demonstrates that radiolytic H₂ can sustain a chemolithotrophic
community; it is not the calibration of the power level, which is why
the cell densities here are anchored to the better-quantified
subseafloor sediment data rather than to the Witwatersrand system. The
modeled RHZ predicts energy-limited subsurface habitability on Mars, if
the required retention conditions are met.

\subsection{3.6 Brine and Hydrated-Film
Sensitivities}\label{brine-and-hydrated-film-sensitivities}

Brine cases retain greater direct oxidant capacity and lie closer to
donor/acceptor balance than the corresponding hydrated-film cases, but
their redox power is approximately an order of magnitude lower because
present-day shallow brines are assigned reduced water activity and
shorter residence. A coupled residence--water-activity sweep shows that
brines do not approach the hydrated-film protected redox power within
the plotted present-day transient-brine window (\(f = 10^{-3}\) to
\(10^{-1}\); \(a_w = 0.40\) to 0.75).

The hydrated-film cases are optimistic transient-film endmembers. When
hydrated-film accessibility is reduced to \(a_w = 0.40\) and
residence/accessibility to 0.10, protected redox power falls to 0.982,
0.845, and 0.559 µJ kg⁻¹ yr⁻¹ for Oxia, Arabia/Mawrth, and southern
Utopia, overlapping the brine cases. The hydrated/brine contrast in
Tables 1 and 2 is therefore best read as a sensitivity to transient-film
accessibility rather than as direct evidence that habitable interlayer
water occurs in any of the three terrains. Hydrated-film accessibility
is, after H₂ retention, a major sensitivity for the active outcome, and
it affects whether modeled redox power remains above the operational
threshold.

\subsection{3.7 Robustness and Comparison with Deep
Radiolysis}\label{robustness-and-comparison-with-deep-radiolysis}

Tier 2 residual-radical closure preserves the active/inactive
classification, the protected depths, the dominant reaction, the
limiting factor, and the dominant uncertainty control across all seven
cases, and reproduces Tier 1 selected medians and intervals to tabulated
precision. Tier 1 is retained as the reference configuration, with Tier
2 as a validated robustness check.

Converted to volumetric production, the active shallow GCR cases yield
78--156 nmol H₂ m⁻³ yr⁻¹ over the full 0--2 m column and 73--143 nmol H₂
m⁻³ yr⁻¹ over the protected 10--200 cm interval. These values lie within
the deep basaltic comparison band of Tarnas et al.~(2021), 25--250 nmol
m⁻³ yr⁻¹ after conversion over a 2 m column. The two radiolysis settings
are complementary rather than directly comparable: Tarnas et al.~address
deep, U/Th/K-driven groundwater radiolysis, whereas the present model
addresses shallow, mission-accessible GCR radiolysis in transient
hydrated phases and brines. That two unrelated radiation
sources operating at very different depths produce comparable H₂
magnitudes is a consistency check on the RHZ framework.

\section{4. Discussion}\label{discussion}

\subsection{4.1 Hydrogen Retention and Supported Cell
Density}\label{hydrogen-retention-and-supported-cell-density}

The two principal results combine hydrogen retention with
maintenance-power constraints. In the modeled cases, the state of a
terrain depends strongly on where the radiolytic hydrogen goes, and the
biological interpretation of that state depends on the minimum power
required for cellular maintenance.

Where hydrogen is retained on mineral surfaces or in clay interlayers,
the active terrains deliver a few × 10⁻¹³ W kg⁻¹ of bulk regolith, which
supports at most order 10³ to 10⁵ cells cm⁻³ at subseafloor maintenance
powers and of order one cell cm⁻³ or fewer at culture-based maintenance
rates.
Where hydrogen escapes as connected-pore free gas, the same terrains
fall by two to four orders of magnitude and drop below the modeled
active range at protected depth. The retained-H₂ case is already an
energy-limited system; the connected-pore free-gas case is below the
modeled habitability threshold. The calculation therefore yields the
energy available for maintenance, an upper bound rather than an estimate
of standing biomass or cell abundance.

The framing also explains why the strong donor abundances in Table 1
coexist with the low maximum supported cell densities in Table 3. Radiolysis can produce measurable hydrogen while still
supplying little biological power, because the conversion from a
per-kilogram energy flux to a per-cell power budget is stringent. The
modeled RHZ should therefore be interpreted as a narrow,
condition-dependent energetic regime within the shallow subsurface,
rather than as evidence for a volumetrically extensive or densely
populated biosphere.

\subsection{4.2 The Diagnostic Measurement: Stepped Evolved-Gas Analysis
of
H₂}\label{the-diagnostic-measurement-stepped-evolved-gas-analysis-of-hux2082}

Because the active or inactive model state is sensitive to H₂ retention
and loss, and because the two H₂-retention states leave distinct
thermal-release fingerprints, evolved-gas analysis can discriminate
between them. The temperatures in this section are instrumental heating
temperatures during stepped EGA, not ambient martian environmental
temperatures. Connected-pore free gas should evolve rapidly during
low-temperature heating (\textless150 °C) with no high-temperature
persistence. Sorbed or mineral-associated H₂, bound on clay surfaces, in
smectite interlayers, or on Fe-bearing oxides, should release at
intermediate heating steps (approximately 200--400 °C) before
mineralogical decomposition becomes significant. Above approximately 400
°C, structural H release from Fe²⁺-bearing phyllosilicates by oxidative
dehydrogenation produces an H₂ signal that is mineralogically diagnostic
but not, by itself, indicative of radiolytic retention (Lempart et al.,
2020). Discrimination of radiolytic from mineralogical H₂ therefore
requires the full release-temperature distribution, not the integrated
H₂ yield, together with a careful correction for the QMS
water-fragmentation contribution at low temperature.

Stepped or staircase evolved-gas analysis (EGA) of H₂ from a
protected-depth sample is therefore the most direct measurement for
testing the H₂-retention component of the modeled RHZ. Sutter et
al.~(2017) have demonstrated this approach in situ using SAM on
Curiosity, with stepped pyrolysis to approximately 870 °C across
multiple Gale samples and direct mass-spectrometric detection of H₂ at
m/z = 2. The published SAM record at Gale is dominated by the
high-temperature mineralogical signature, with no resolvable radiolytic
excess at low-to-intermediate temperatures (Section 3.2). Application of
the same methodology at any active terrain would directly test the
prediction.

\subsection{4.3 Implications for ExoMars Rosalind Franklin and
Tianwen-3}\label{implications-for-rosalind-franklin-and-tianwen-3}

The two missions positioned to test the RHZ are ExoMars Rosalind Franklin at
Oxia Planum and Tianwen-3 returning material from southern Utopia.

For ExoMars Rosalind Franklin, the 2 m drill provides direct access to the
protected-depth interval in which the model predicts non-zero radiolytic
redox power. A stepped pyrolysis sequence with H₂ monitoring at m/z = 2
on a clay-bearing subsurface sample, paired with hydration-state
mineralogy, would help attribute the H₂-release distribution to specific
mineral hosts. MOMA's pyrolysis ovens reach approximately 850 °C with
stepped-temperature sequences (Goesmann et al., 2017), so the underlying
thermal capability is comparable to that of SAM. MOMA is designed primarily for
organic-molecule characterization, and quantitative H₂ release is not in
its published baseline product set; H₂ detection during stepped
pyrolysis would require an intentionally selected measurement mode,
primarily through the GC thermal-conductivity-detector path or a
targeted ion-trap configuration at low m/z. Such an observation would be
an operationally specific test of the RHZ prediction rather than a
routine mission data product.

For Tianwen-3 returning southern Utopia material, and for Mars Sample
Return analogs in the Arabia/Mawrth window, returned-sample analysis
can apply the same test without requiring changes to a flight
instrument. Low-temperature stepped pyrolysis with quantitative gas
chromatography of evolved H₂ sits within the standard sample-handling
envelope. For the modeled RHZ, returned-sample stepped EGA would
establish on a per-sample basis whether shallow radiolytic activity is a
credible component of the local biogeochemistry. Other measurements
(DSC/TGA water-reservoir partitioning, ion chromatography for
oxychlorines, Fe-redox speciation, and Raman/IR mineral-hydration
context) are necessary to characterize the partitioning of water
available to radiolysis and the acceptor mineralogy, but they are not,
individually, sufficient to determine H₂ retention and loss.

\subsection{4.4 Habitability and Biosignature
Preservation}\label{habitability-and-biosignature-preservation}

The radiation-driven preservation factor is target dependent and should
not be conflated with habitability. A dose scale of \(D_{37} = 10^6\) Gy
is best treated as a microbial radiation-tolerance stress case; it is
not an appropriate proxy for long-term preservation of amino acids or
other organic biosignatures. \(D_{37} = 10^5\) Gy is the appropriate
conservative simple-organic/biosignature sensitivity (Pavlov et al.,
2022). At protected depths, the 10 Myr \(D_{37} = 10^5\) Gy preservation
factors are \(2.6 \times 10^{-4}\) to \(4.2 \times 10^{-4}\) for the
active cases, against 0.44--0.46 at \(D_{37} = 10^6\) Gy. The active
terrains identified here are better described as targets for testing
contemporary or geologically recent radiolytic energy availability, not
as generic organic-preservation targets. Preservation of organic
biosignatures over 10 Myr at these depths is severely constrained even
in the active cases. The maximum supported cell density of Section 3.5
and this preservation constraint are consistent: a low-density,
energy-limited community would be expected to leave a low-abundance and
radiation-altered organic record, which is itself a prediction for the
returned-sample search.

\subsection{4.5 Limitations and
Falsifiability}\label{limitations-and-falsifiability}

The results above carry three principal limitations that define the
strongest tests of the model.

First, the supported cell-density estimate is dominated by the choice of
per-cell maintenance power, which spans roughly five orders of magnitude
in the literature (Section 2.4). The cell densities in Table 3
are therefore order-of-magnitude energetic maxima conditioned on
\(P_{\mathrm{maint}}\), not predictions of cell abundance. Across the
entire \(P_{\mathrm{maint}}\) range, the qualitative conclusion is
that the active RHZ is an energy-limited system, of order one cell cm⁻³
or fewer at
culture-based maintenance rates and sparsely populated only at
subseafloor maintenance rates. The absolute number requires an
independent constraint on martian subsurface maintenance energetics that
does not yet exist. On the supply side, the conversion treats radiolytic
redox power as the in situ catabolic free-energy flux (Section 2.4):
\(\Delta G\) is evaluated as \(\Delta G^\circ + RT \ln Q\) at
Mars-conditioned temperature (210 K), scenario-derived water activity
and pH, effective ionic strength, and representative dissolved-species
activities. The distinction is immaterial for the H₂ + H₂O₂ pairing
dominating every active case, because the in situ value at default Mars
conditions is approximately \(-310\ \mathrm{kJ\ mol^{-1}}\), more than
\(15\times\) the biological energy quantum. It would matter, however,
for any terrain whose budget rested on a weaker acceptor closer to the
biological energy quantum.

Second, the active predictions rest on the reference H₂ escape
coefficient, a plausibility-based effective first-order loss rate
rather than a measurement (Section 2.5). This is the largest sampled
uncertainty control in every active case, and the model output is most
sensitive to the variable that the diagnostic measurement of Section 4.2
directly constrains. The framework therefore treats H₂ retention and
loss as a testable sensitivity rather than as a fixed environmental
property.

Third, the three active terrains cannot be ranked against one another on
radiolysis alone. The paired shared-prior analysis preserves the
ordering of median redox power, but the independent 5--95\% intervals
overlap strongly (Section 3.3), and the paired diagnostic excludes
terrain-specific systematic errors in the accessibility of water
available to radiolysis, sorption state, and acceptor mineralogy.
Radiolytic redox power is a within-framework activity metric, not a
site-prioritization metric, and it should not be used to choose between
Oxia, southern Utopia, and Arabia/Mawrth.

The active predictions are conditional on three falsifiable
requirements: a transient hydrated-film state with water available to
radiolysis at protected depths; sorbed or mineral-associated H₂
retention; and accessible electron acceptors at concentrations
consistent with the assumed mineralogy. Each is independently testable.
Predictions for Oxia, southern Utopia, and Arabia/Mawrth would be
weakened by protected-depth samples showing a low fraction of water
available to radiolysis, absent retained oxidants or oxychlorine
products, no H₂-bearing redox products above the mineralogical and
instrumental backgrounds quantified in Section 4.2, or mineralogy
inconsistent with accessible Fe(III), sulfate, nitrate, or oxychlorine
acceptors. The modeled RHZ should therefore be evaluated against
forthcoming in situ and returned-sample measurements.

\section{5. Conclusions}\label{conclusions}

Shallow martian radiolysis is a viable energy source for
chemolithotrophic life only under a specific and testable set of
physical--chemical conditions, and even where those conditions are met
the resulting biosphere would be energy limited. The modeled state of a
terrain is most sensitive to the retention and loss of radiolytic
hydrogen: under sorbed or mineral-associated retention, the active
terrains of Oxia Planum, Arabia/Mawrth, and southern Utopia deliver a
few × 10⁻¹³ W kg⁻¹ of bulk regolith at protected depth; under
connected-pore free-gas retention the same terrains fall by two to four
orders of magnitude and drop below the active range. Converted to
biology, even the retained-H₂ case supports at most order 10³ to 10⁵
cells cm⁻³, and only at the per-cell power of Earth's energy-limited
subseafloor biosphere, placing the modeled RHZ closer to systems
sustained by radiolytic hydrogen in Earth's deep subsurface than to a
high-cell-density biosphere. Gale Crater, used as a
Curiosity/DAN-constrained reference case, is radiolytically inactive at
protected depth under both H₂-retention assumptions, and the published
SAM evolved-gas analyses are consistent with that null.

The modeled RHZ therefore depends on more than whether radiolytic energy
is produced. It depends on whether the host material retains radiolytic
hydrogen on biologically relevant timescales, and on whether the
resulting energy flux can support detectable cell densities. Both
questions are testable with stepped evolved-gas analysis of H₂ from a
protected-depth sample: demonstrated in situ at Gale by SAM, within
reach of MOMA on ExoMars Rosalind Franklin with operational adaptation, and
applicable to Tianwen-3 and Mars Sample Return material. The relevant
measurement pathways are represented in current or planned mission
architectures.

\section{6. Figures}\label{figures}

\clearpage

\begin{figure}[p]
\centering
\includegraphics[width=\textwidth]{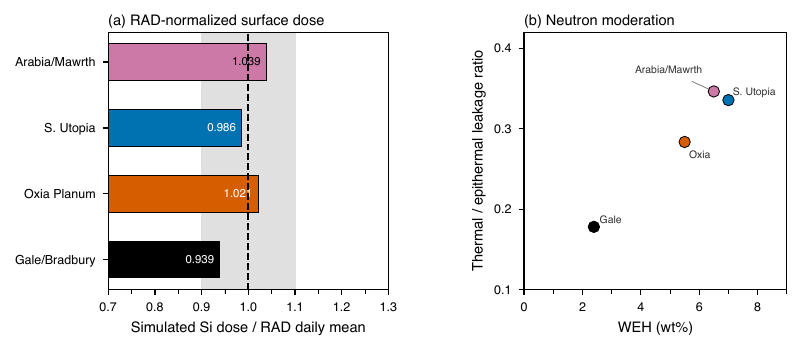}
\caption{Transport validation observables. (a) Simulated daily mean dose in the surface silicon slab, normalized to the adopted Curiosity/RAD reference. Values near unity indicate that the transport normalization is consistent with landed dosimetry at the level needed for the terrain-family comparison; the panel is not used to rank habitability. (b) Upward thermal-to-epithermal neutron leakage ratio at the top of the regolith as a qualitative moderation diagnostic against water-equivalent hydrogen. The ratio measures neutron moderation by subsurface hydrogen, qualitatively analogous to the quantity probed by DAN, and is not a simulated DAN count rate.}
\label{fig:validation}
\end{figure}

\begin{figure}[p]
\centering
\includegraphics[width=\textwidth]{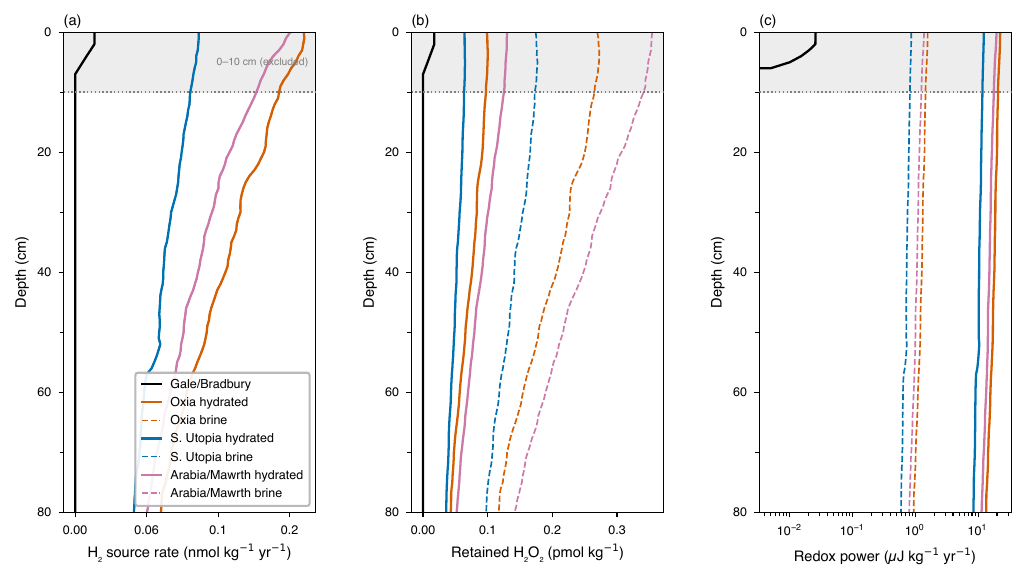}
\caption{Depth-resolved radiolysis products by terrain. (a) H$_2$ source rate, (b) retained H$_2$O$_2$, and (c) radiolytic redox power, for the seven canonical cases. Profiles are shown as a 5~cm centered moving average. The shaded 0--10 cm interval is excluded from the protected-depth interpretation.}
\label{fig:depth-profiles}
\end{figure}

\begin{figure}[p]
\centering
\includegraphics[width=\textwidth]{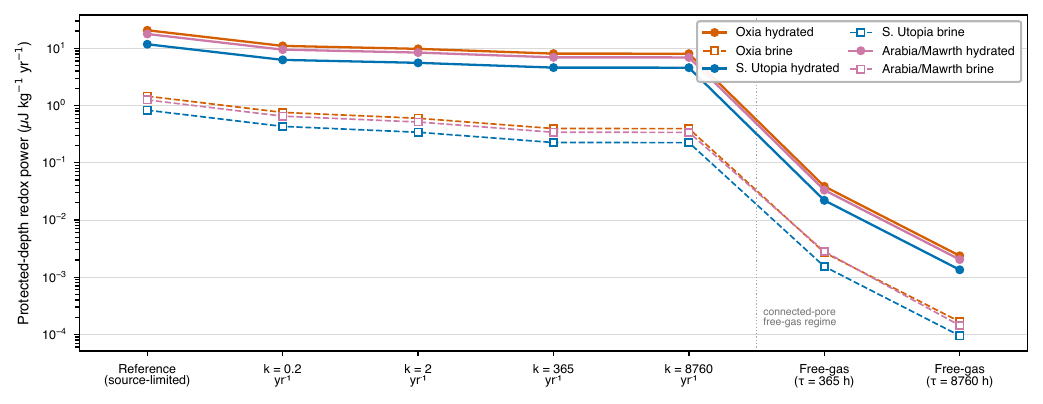}
\caption{Sensitivity to H$_2$ retention and loss. Protected-depth radiolytic redox power across the sorbed baseline, intermediate gas-loss sensitivity cases, and the connected-pore free-gas case. Redox power decreases by approximately 285--4700$\times$ between the sorbed baseline and the connected-pore free-gas case.}
\label{fig:h2-retention}
\end{figure}

\begin{figure}[p]
\centering
\includegraphics[width=\textwidth]{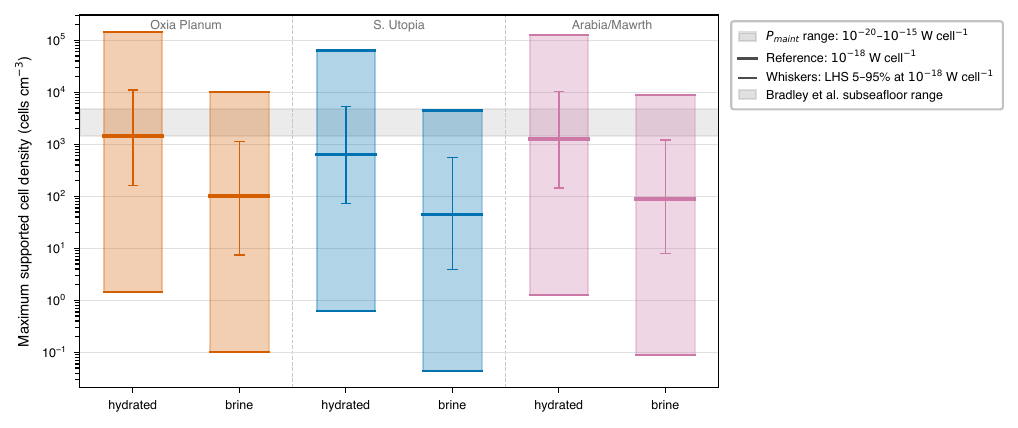}
\caption{Maintenance-power conversion to supported cell density. Protected-depth power density for the active terrains converted to maximum supported cell density across the maintenance-power range (10$^{-15}$ to 10$^{-20}$ W per cell). Vertical whiskers show the 5--95\% Latin-hypercube uncertainty interval in radiolytic redox power, propagated at 10$^{-18}$ W per cell. The grey horizontal band shows the cell densities implied by the Bradley et al. (2020) subseafloor cell-specific power range for the Oxia hydrated reference case.}
\label{fig:cell-density}
\end{figure}

\clearpage

\section{7. Acknowledgements}\label{acknowledgements}

The author thanks the Geant4 Collaboration and CERN for the development
and maintenance of the Geant4 simulation toolkit. The author
acknowledges the NASA Mars Science Laboratory mission and its RAD, DAN,
and SAM instrument teams, together with the NASA Planetary Data System,
for the landed radiation, neutron, and evolved-gas datasets that provide
essential context for model validation and the Gale Crater comparison.
The author also acknowledges the European Space Agency ExoMars programme
and the China National Space Administration Tianwen-3 mission teams,
whose landing-site characterizations motivate the terrains examined
here. Computational resources were provided by the NYUAD
High-Performance Computing facility.

\section{8. Author Contributions}\label{author-contributions}

D.A. is the sole author and was responsible for the conceptualization,
methodology, software, formal analysis, investigation, visualization,
and writing of this study.

\section{9. Statements and
Declarations}\label{statements-and-declarations}

\subsection{9.1 Ethical Approval}\label{ethical-approval}

This computational study did not involve human participants, human or
animal subjects, human tissue, or fieldwork; ethical approval was not
required.

\subsection{9.2 Declaration of Conflicting
Interests}\label{declaration-of-conflicting-interests}

The author declares no conflict of interest.

\subsection{9.3 Funding}\label{funding}

This work was supported by NYUAD Research Institute grant CG 014.

\subsection{9.4 Data and Code
Availability}\label{data-and-code-availability}

The processed data products underlying the figures and tables (the
depth-resolved radiolysis and redox-power profiles, the protected-depth
uncertainty summaries, the supported-cell-density conversions, and the
H₂-retention and accessibility sensitivity-test outputs) will be
deposited in a public repository upon acceptance. The transport stage uses the publicly
available Geant4 toolkit (Agostinelli et al., 2003; Allison et al.,
2006; Allison et al., 2016); the custom radiolysis-chemistry,
redox-power, and cell-density code, together with further model outputs,
is available from the author on reasonable request. The Methods
(Section 2) specify the governing physics, chemistry, source terms, and
parameter values in full, so that the model can be reproduced
independently.

\section{References}\label{references}
\setlength{\parindent}{0pt}
\setlength{\parskip}{0.6em}

Agostinelli S, Allison J, Amako K, Apostolakis J, Araujo H, Arce P et al. (2003) Geant4: a simulation toolkit. Nuclear Instruments and Methods in Physics Research A 506: 250--303. DOI: 10.1016/S0168-9002(03)01368-8

Allison J, Amako K, Apostolakis J, Araujo H, Arce Dubois P, Asai M et al. (2006) Geant4 developments and applications. IEEE Transactions on Nuclear Science 53: 270--278. DOI: 10.1109/TNS.2006.869826

Allison J, Amako K, Apostolakis J, Arce P, Asai M, Aso T et al. (2016) Recent developments in Geant4. Nuclear Instruments and Methods in Physics Research A 835: 186--225. DOI: 10.1016/j.nima.2016.06.125

Atri D (2016) On the possibility of galactic cosmic ray-induced radiolysis-powered life in subsurface environments in the Universe. Journal of the Royal Society Interface 13: 20160459. DOI: 10.1098/rsif.2016.0459

Atri D (2020) Investigating the biological potential of galactic cosmic ray-induced radiation-driven chemical disequilibrium in the Martian subsurface environment. Scientific Reports 10: 11646. DOI: 10.1038/s41598-020-68715-7

Atri D, Kamenetskiy M, May M, Kalra A, Castelblanco A and Quiñones-Camacho A (2025) Estimating the potential of ionizing-radiation-induced radiolysis for microbial metabolism on terrestrial planets and satellites with rarefied atmospheres. International Journal of Astrobiology 24: e9. DOI: 10.1017/S1473550425100025

Bishop JL, Dobrea EZN, McKeown NK, Parente M, Ehlmann BL, Michalski JR, Milliken RE, Poulet F, Swayze GA, Mustard JF, Murchie SL and Bibring JP (2008) Phyllosilicate diversity and past aqueous activity revealed at Mawrth Vallis, Mars. Science 321: 830--833. DOI: 10.1126/science.1159699

Bradley JA, Arndt S, Amend JP, Burwicz E, Dale AW, Egger M and LaRowe DE (2020) Widespread energy limitation to life in global subseafloor sediments. Science Advances 6: eaba0697. DOI: 10.1126/sciadv.aba0697

Chivian D, Brodie EL, Alm EJ, Culley DE, Dehal PS, DeSantis TZ, Gihring TM, Lapidus A, Lin LH, Lowry SR, Moser DP, Richardson PM, Southam G, Wanger G, Pratt LM, Andersen GL, Hazen TC, Brockman FJ, Arkin AP and Onstott TC (2008) Environmental genomics reveals a single-species ecosystem deep within Earth. Science 322: 275--278. DOI: 10.1126/science.1155495

Dartnell LR, Desorgher L, Ward JM and Coates AJ (2007) Modelling the surface and subsurface martian radiation environment: implications for astrobiology. Geophysical Research Letters 34: L02207. DOI: 10.1029/2006GL027494

Davila AF and Schulze-Makuch D (2016) The last possible outposts for life on Mars. Astrobiology 16: 159--168. DOI: 10.1089/ast.2015.1380

Dzaugis M, Spivack AJ and D'Hondt S (2018) Radiolytic H₂ production in Martian environments. Astrobiology 18: 1137--1146. DOI: 10.1089/ast.2017.1654

Gleeson LJ and Axford WI (1968) Solar modulation of galactic cosmic rays. Astrophysical Journal 154: 1011--1026. DOI: 10.1086/149822

Goesmann F, Brinckerhoff WB, Raulin F, Goetz W, Danell RM, Getty SA et al. (2017) The Mars Organic Molecule Analyzer (MOMA) instrument: characterization of organic material in martian sediments. Astrobiology 17: 655--685. DOI: 10.1089/ast.2016.1551

Guo J, Zeitlin C, Wimmer-Schweingruber RF, Rafkin S, Hassler DM, Posner A, Heber B, Köhler J, Ehresmann B, Appel JK, Böhm E, Böttcher S, Burmeister S, Brinza DE, Lohf H, Martin C, Kahanpää H and Reitz G (2015) Modeling the variations of dose rate measured by RAD during the first MSL martian year: 2012--2014. Astrophysical Journal 810: 24. DOI: 10.1088/0004-637X/810/1/24

Hassler DM, Zeitlin C, Wimmer-Schweingruber RF, Ehresmann B, Rafkin S, Eigenbrode JL et al. (2014) Mars' surface radiation environment measured with the Mars Science Laboratory's Curiosity rover. Science 343: 1244797. DOI: 10.1126/science.1244797

Hoehler TM and Jørgensen BB (2013) Microbial life under extreme energy limitation. Nature Reviews Microbiology 11: 83--94. DOI: 10.1038/nrmicro2939

Hou Z, Liu J, Pang F, Wang Y, Li Y, Xu M et al. (2025) In search of signs of life on Mars with China's sample return mission Tianwen-3. Nature Astronomy 9: 783--792. DOI: 10.1038/s41550-025-02572-0

LaRowe DE and Amend JP (2015) Power limits for microbial life. Frontiers in Microbiology 6: 718. DOI: 10.3389/fmicb.2015.00718

Lasne J, Noblet A, Szopa C, Navarro-González R, Cabane M, Poch O, Stalport F, François P, Atreya SK and Coll P (2016) Oxidants at the surface of Mars: a review in light of recent exploration results. Astrobiology 16: 977--996. DOI: 10.1089/ast.2016.1502

LaVerne JA (2000) Track effects of heavy ions in liquid water. Radiation Research 153: 487--496. DOI: 10.1667/0033-7587(2000)153[0487:TEOHII]2.0.CO;2

Lempart M, Derkowski A, Strączek T and Kapusta C (2020) Systematics of H₂ and H₂O evolved from chlorites during oxidative dehydrogenation. American Mineralogist 105: 932--944. DOI: 10.2138/am-2020-7326

Lin LH, Wang PL, Rumble D, Lippmann-Pipke J, Boice E, Pratt LM, Lollar BS, Brodie EL, Hazen TC, Andersen GL, DeSantis TZ, Moser DP, Kershaw D and Onstott TC (2006) Long-term sustainability of a high-energy, low-diversity crustal biome. Science 314: 479--482. DOI: 10.1126/science.1127376

Mitrofanov IG, Nikiforov SY, Djachkova MV, Lisov DI, Litvak ML, Sanin AB and Vasavada AR (2022) Water and chlorine in the martian subsurface along the traverse of NASA's Curiosity rover: 1. DAN measurement profiles along the traverse. Journal of Geophysical Research: Planets 127: e2022JE007327. DOI: 10.1029/2022JE007327

Pastina B and LaVerne JA (2001) Effect of molecular hydrogen on hydrogen peroxide in water radiolysis. Journal of Physical Chemistry A 105: 9316--9322. DOI: 10.1021/jp012245j

Pavlov AA, Vasilyev G, Ostryakov VM, Pavlov AK and Mahaffy P (2012) Degradation of the organic molecules in the shallow subsurface of Mars due to irradiation by cosmic rays. Geophysical Research Letters 39: L13202. DOI: 10.1029/2012GL052166

Pavlov AA, McLain HL, Glavin DP, Roussel A, Dworkin JP, Elsila JE and Yocum KM (2022) Rapid radiolytic degradation of amino acids in the martian shallow subsurface: implications for the search for extinct life. Astrobiology 22: 1099--1115. DOI: 10.1089/ast.2021.0166

Poulet F, Bibring JP, Mustard JF, Gendrin A, Mangold N, Langevin Y, Arvidson RE, Gondet B, Gomez C and the OMEGA Team (2005) Phyllosilicates on Mars and implications for early martian climate. Nature 438: 623--627. DOI: 10.1038/nature04274

Quantin-Nataf C, Carter J, Mandon L, Thollot P, Balme M, Volat M et al. (2021) Oxia Planum: the landing site for the ExoMars Rosalind Franklin rover mission: geological context and prelanding interpretation. Astrobiology 21: 345--366. DOI: 10.1089/ast.2019.2191

Quinn RC, Martucci HFH, Miller SR, Bryson CE, Grunthaner FJ and Grunthaner PJ (2013) Perchlorate radiolysis on Mars and the origin of martian soil reactivity. Astrobiology 13: 515--520. DOI: 10.1089/ast.2013.0999

Sutter B, McAdam AC, Mahaffy PR, Ming DW, Edgett KS, Rampe EB et al. (2017) Evolved gas analyses of sedimentary rocks and eolian sediment in Gale Crater, Mars: results of the Curiosity rover's Sample Analysis at Mars instrument from Yellowknife Bay to the Namib Dune. Journal of Geophysical Research: Planets 122: 2574--2609. DOI: 10.1002/2016JE005225

Tarnas JD, Mustard JF, Sherwood Lollar B, Stamenković V, Cannon KM, Lorand JP, Onstott TC, Michalski JR, Warr O, Palumbo AM and Plesa AC (2021) Earth-like habitable environments in the subsurface of Mars. Astrobiology 21: 741--756. DOI: 10.1089/ast.2020.2386

Tijhuis L, van Loosdrecht MCM and Heijnen JJ (1993) A thermodynamically based correlation for maintenance Gibbs energy requirements in aerobic and anaerobic chemotrophic growth. Biotechnology and Bioengineering 42: 509--519. DOI: 10.1002/bit.260420415

Usoskin IG, Alanko-Huotari K, Kovaltsov GA and Mursula K (2005) Heliospheric modulation of cosmic rays: monthly reconstruction for 1951--2004. Journal of Geophysical Research 110: A12108. DOI: 10.1029/2005JA011250

Vago JL, Westall F, Coates AJ, Jaumann R, Korablev O, Ciarletti V et al. (2017) Habitability on early Mars and the search for biosignatures with the ExoMars rover. Astrobiology 17: 471--510. DOI: 10.1089/ast.2016.1533

\end{document}